\documentstyle[11pt,newpasp,twoside,epsf]{article}
\def\edcomment#1{\iffalse\marginpar{\raggedright\sl#1\/}\else\relax\fi}
\reversemarginpar

\begin{document}
\title{Low Frequency Radio Emission of Pulsar PSR J1907+0919 Associated with the Magnetar SGR 1900+14}
\author{Yu. P. Shitov, V. D. Pugachev \& S. M. Kutuzov}
\affil{Pushchino Radioastronomy Observatory, Astro Space Center of Lebedev
Physical Institute, Pushchino, 142292, Russia, shitov@prao.psn.ru}

\begin{abstract}
The soft gamma repeater SGR 1900+14 was observed in Pushchino
observatory since 1988 December using BSA radio telescope
operating at 111 MHz. We have detected the pulsed radio emission
(Shitov 1999) with the same 5.16 s period that was reported
earlier for this object (Hurley et al. 1998). The timing analysis
has shown that this new radio pulsar PSR J1907+0919 associated
with SGR 1900+14 has a superstrong magnetic field, which is $8.1
\cdot 10^{14}$G, thereby confirming that it is a "magnetar"
(Duncan \& Thompson 1992; Kouveliotou et al. 1999). The dispersion
measure of PSR J1907+0919 is $281.4(9) pc \cdot cm^{-3}$ which
gives an estimate of the pulsar's distance as about 5.8 kpc.
\end{abstract}

\section{Introduction}

Among known soft gamma- ray repeaters to date, only SGR 1900+14
and SGR 1806-20 are the objects for which a secular spin-down of
the pulse periods (5.16 s and 7.47 s accordingly) with $\dot{P}$
of order $10^{-10}$s/s was detected and thereby was established
that these SGRs are neutron stars with a superstrong magnetic
field of order $10^{15}$G (Kouveliotou et al. 1998; Kouveliotou et
al. 1999), called as a  "magnetars" (Duncan \& Thompson 1992).

Since the end of 1998 we carried out the observations of the SGR
1900+14 at low frequency (111 MHz) and have detected the periodic
pulsed radio emission from this magnetar (Shitov 1999). In this
paper we report the results of our observations obtained till
August 1999.

\section{Observations}

The observations of the magnetar SGR 1900+14 were started since
1998 December using the BSA radio telescope operating now at the
new frequency of 111 MHz. (In 1998 October BSA - wavelength dipole
array with dimensions of 187 x 384 m, transit time of
$200s/Cos(\delta)$ - was reconstructed to shift the previous
operating frequency of 102.5 MHz to the new one of 111 MHz
(Kutuzov et al. 1999) ) We used 128 x 20 kHz filterbank receiver
and multichannel recording system to record the individual pulses
during the BSA transit time. Different sampling intervals of
22.6816 and 20.1728 ms were used. All further processing
procedures: cleaning of the noise signal in a each frequency
channel, dedispersion, spectral analysis, the folding of
dedispersed data for integrated pulse profile and the timing
analysis were made "off line".

\section{The detection of radio pulses from SGR 1900+14}

For the first time we have detected pulsed with 5.161-s period
radio emission from SGR 1900+14 in 1998 December 12, when as
result of dedispersion procedures the integrated pulse profile
with a good enough signal/noise ratio appeared with a dispersion
measure ( DM ) of about $280 pc \cdot cm^{-3}$. It was surprising
that the revealed profile was quite narrow, with the width of
about $7\fdg5$. The next radio detection of SGR 1900+14 like the
first one was only in 1999 January 6 (note that between this dates
the antenna array was often covered by hoar-frost). Fig. 1
demonstrates these two detection of the SGR signals at 111 MHz.

\begin{figure}
\vspace{0cm} \hbox{\hspace{0cm}\epsfxsize=6cm \epsfbox{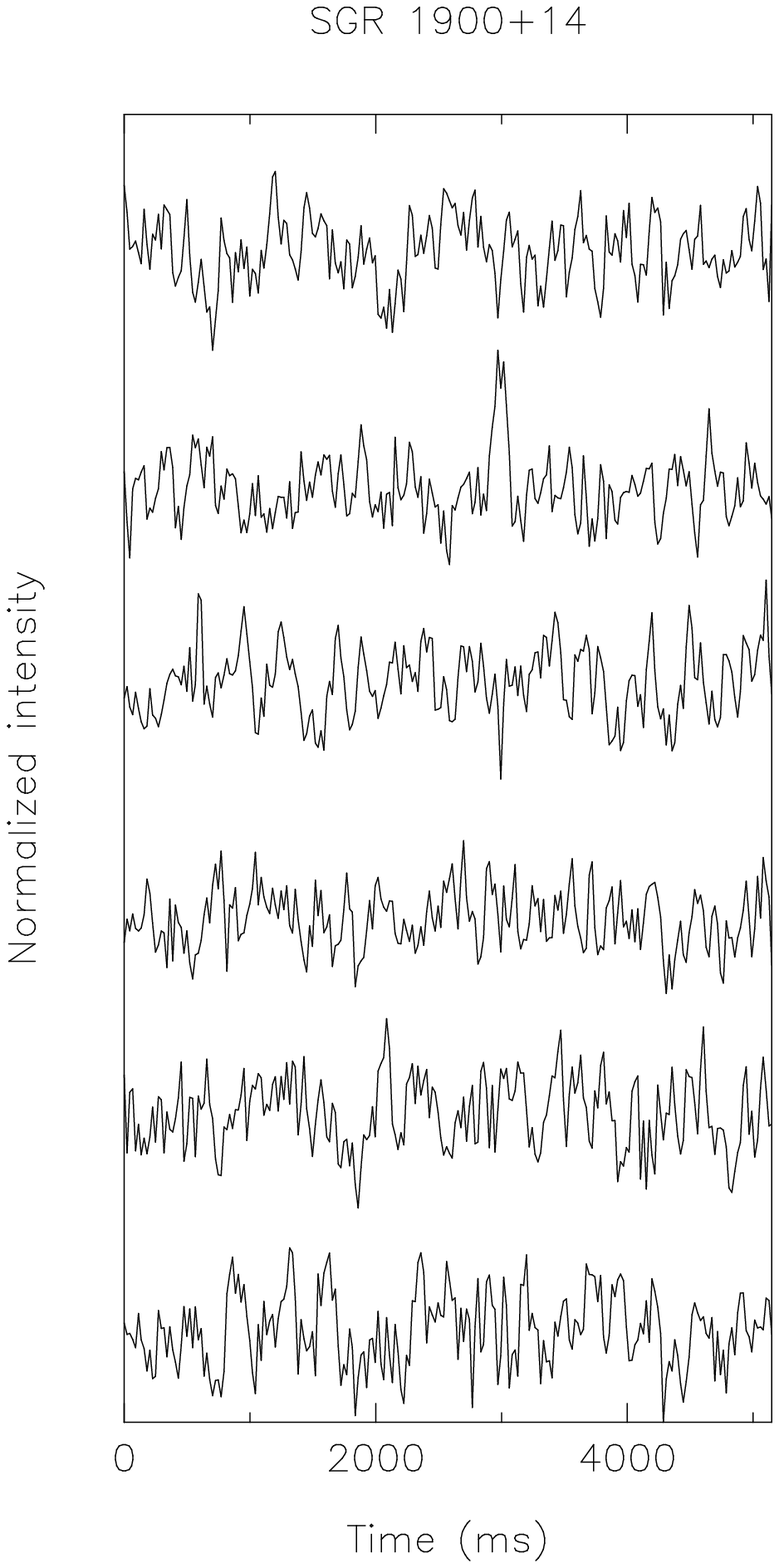}
\epsfxsize=6cm \epsfbox{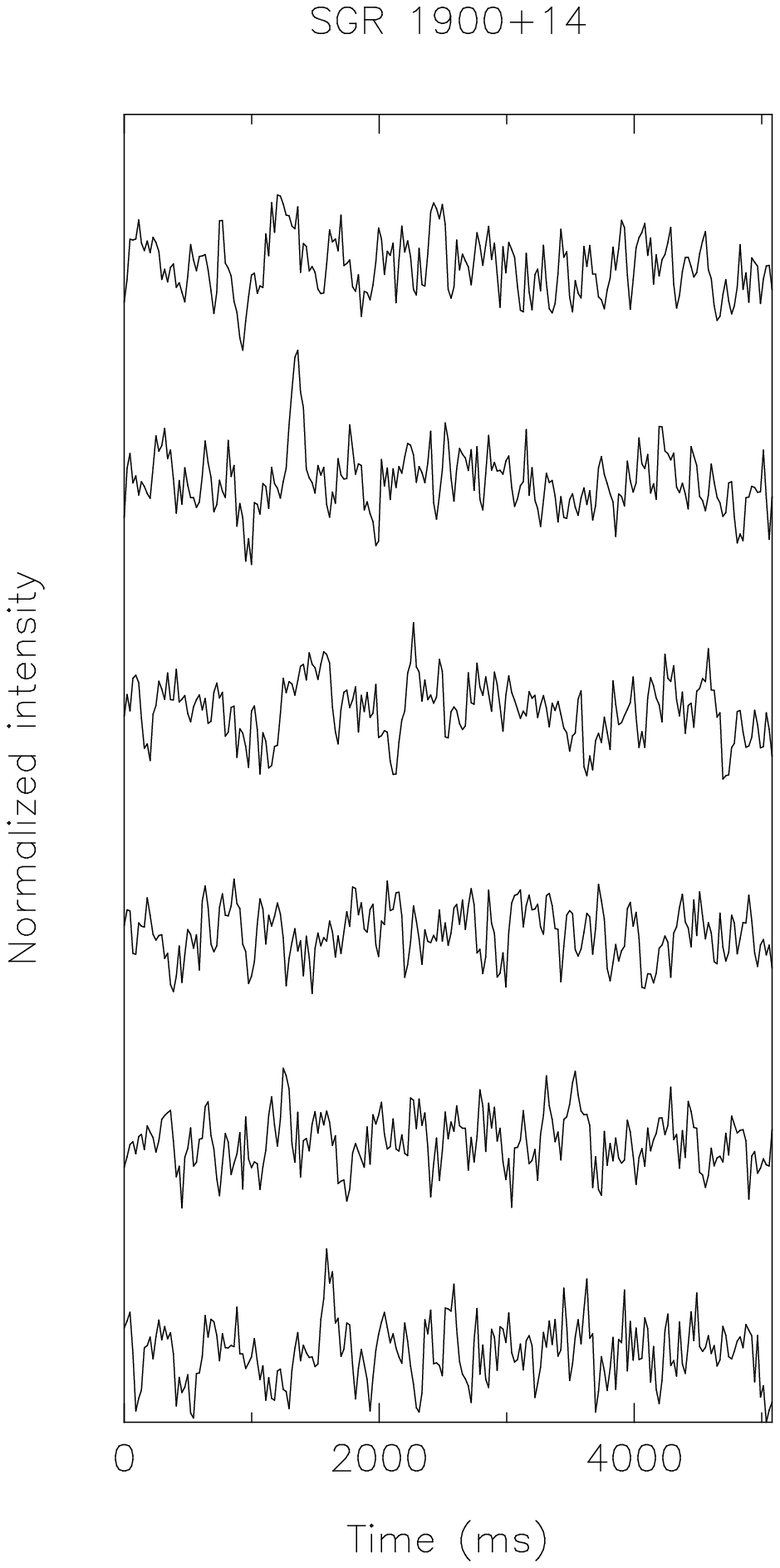}} \vspace{0cm}

\caption{The dedispersed with different DM values and folded with
the 5.161 s period records of SGR 1900+14, obtained at 111 MHz in
1998 December 12 (left) and 1999 January 6 (right). DM = 0
(bottom), 70, 140, 210, 281, 350}
\end{figure}

During subsequent regular observations the pulses from SGR 1900+14
with the same DM values and similar pulse width of about 100 ms
were detected repeatedly. The estimated for the best records mean
flux density at 111 MHz is approximately 50 mJy. The measured
value of $DM = 281.4 (9) pc \cdot cm^{-3}$. So far as the period
derivative $\dot{P}$ was known for the magnetar with the limited
accuracy the pulses were recorded in a unpredictable phases
(Fig.2, left) till the middle of January 1999 when $P$ and
$\dot{P}$ were improved as a result of the timing analysis.

\begin{figure}
\vspace{0cm} \hbox{\hspace{0cm}\epsfxsize=6cm \epsfbox{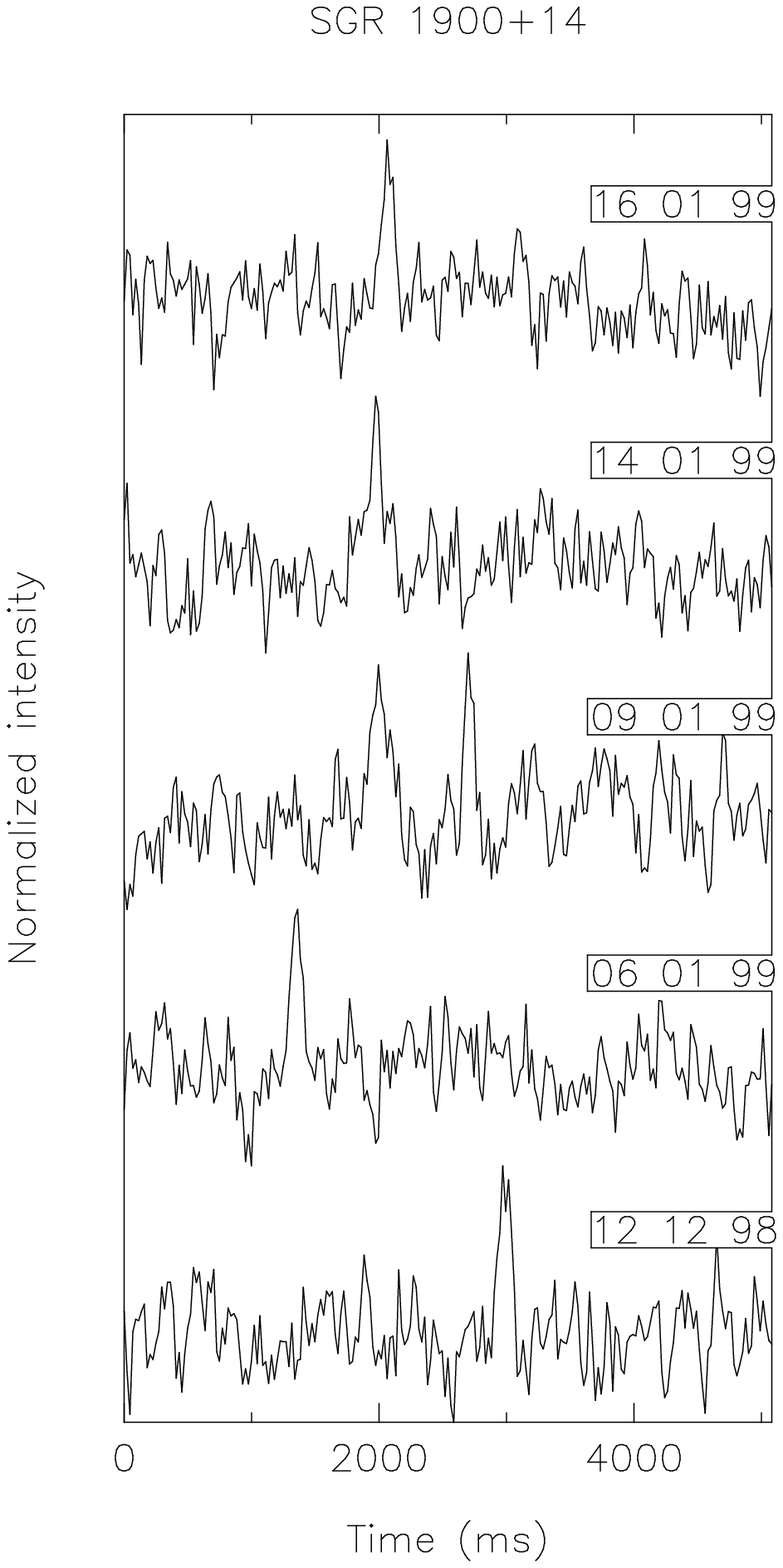}
\epsfxsize=6cm \epsfbox{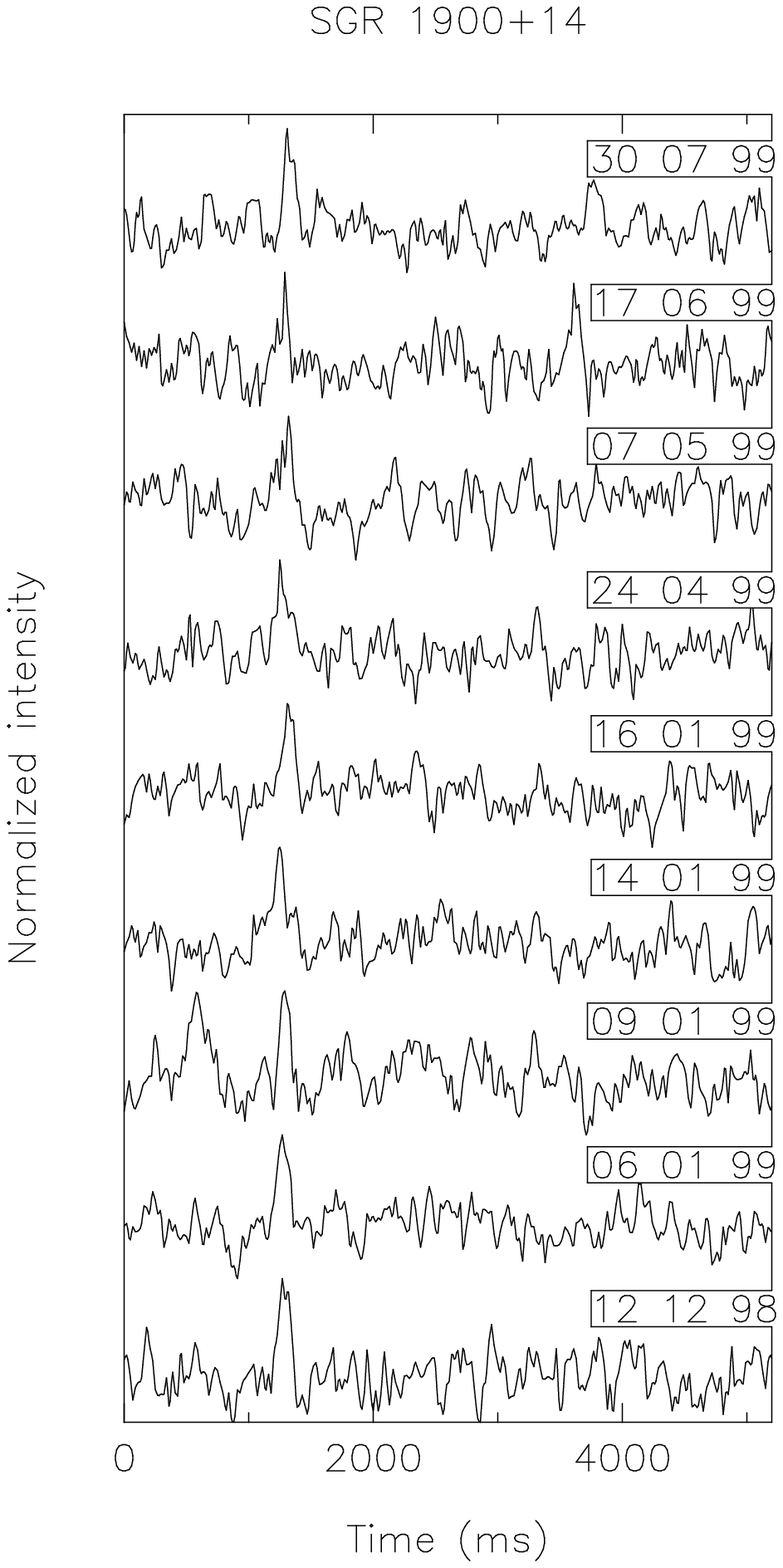}} \vspace{0cm}

\caption{Left - some of the
first records of SGR 1900+14 obtained at 111 MHz. After improving
of the $P$ and $\dot{P}$ values the pulse profiles on January 14
and 16 were observed in the precalculated phases. Right - the time
aligned pulse profiles of SGR 1900+14 barycentric phases of which
are in accordance with the best fitted $P$, $\dot{P}$, and
$\ddot{P}$ values. }
\end{figure}
Fig.2 (right) demonstrates some pulse profiles of  SGR 1900+14,
the phase of which aligned in an accordance with the timing
solution.

There were a number of records in which we have detected
interpulses with amplitude (and, of course, dispersion measure)
like the main pulse, and which were located in different phases.
Two examples of these interpulses are seen in Fig. 2 (09 01 99 and
17 06 99).

\section{Timing analysis}
For timing analysis of our data as the initial parameters of  SGR
1900+14 we used the values of P = 5.160199(2) s, $\dot{P} =
1.14(23) \cdot 10^{-10}$ s/s, MJD = 51056.0, obtained by
Kouveliotou et al. (1999) in 1998 August 28. The exact position of
the magnetar: $R.A. (J2000) = 19h 07m 14\fs33$, $Dec.(J2000) =
09\deg 19\arcmin 20\farcs1$ , obtained with the VLA observations
of an outburst of relativistic particles from SGR 1900+14 (Frail,
Kulkarni \& Bloom 1999) we used in our analysis. The first timing
solution for $P$ and $\dot{P}$ was found for the first records,
obtained in the date interval Dec. 12 1998 - Jan. 14 1999, after
that in the subsequent observations with improved parameters the
pulses were detected in the precalculated phases. In IAUC 7110 we
have reported the solution: P = 5.161297854 (83)s, $\dot{P}$ =
1.23228 (34)s/s, which was found for the interval Dec. 12 1998 -
Feb. 4 1999.

\begin{figure}
\plotfiddle{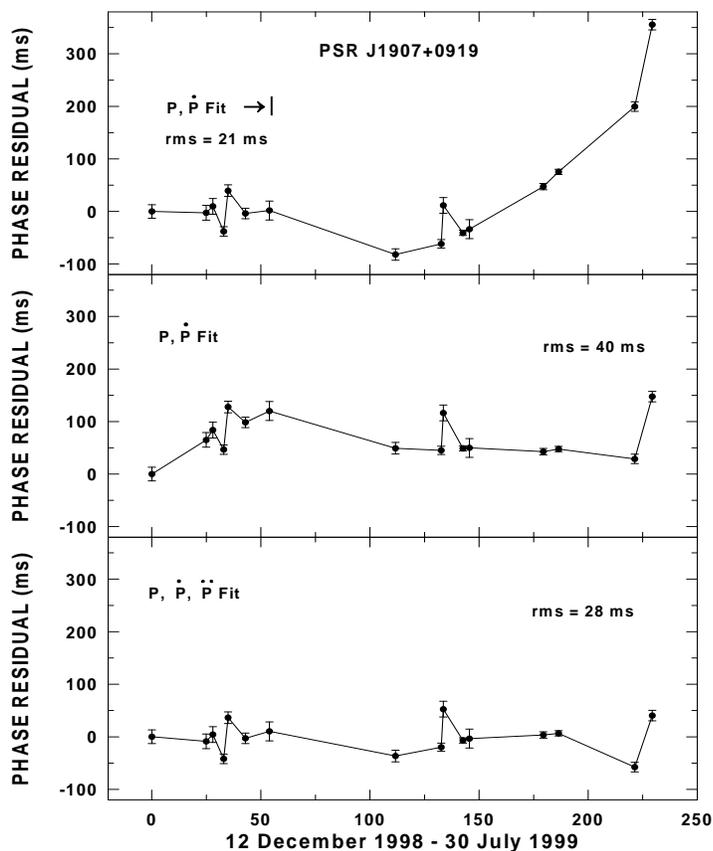}{11.4cm}{0}{55}{50}{-175}{-65}
\caption{Timing residuals for different parameters $P$, $\dot{P}$
and $\ddot{P}$: from IAUC 7110 (top); $\ddot{P}$ = 0.0 (middle);
best fitted (bottom)}
\end{figure}

In this paper we present the results, obtained during Dec. 12 1998
- July 30 1999. We have selected 16 best records and found for
this dates interval the following timing solution: P = 5.161297899
(67) s, $\dot{P} = 1.23197 (15)\cdot 10^{-10}$ s/s, $\ddot{P} =
+0.53 (14) \cdot 10^{-20}$ s/s/s, MJD = 51159.4605 with the RMS of
phase residual of 28 ms. The inclusion of the second period
derivative essentially improved the timing in this interval of
dates (see Fig. 3). For the epoch MJD = 51056.0, found parameters
give P = 5.1601969 s, that is in a good agreement with
Kouveliotou's et al. (1999) measurements.

\section{Conclusions}

    New radio pulsar PSR J1907+0919 associated with the soft gamma
repeater SGR 1900+14 is representative of a new class of pulsars
with a superstrong magnetic field, slow down value of which for
this pulsar is $8.1 \cdot 10^{14}$ G. Presented results confirm
that this object is a magnetar (Duncan \& Thompson 1992;
Kouveliotou et al. 1999). The pulsar distance of  about 5.8 kpc
determined from dispersion measure $DM = 281 pc \cdot cm^{-3}$
supports the suggested earlier in a number of papers genetic
connection of SGR 1900+14 with supernova remnants SNR G42.8+0.6.
As timing analysis have shown there is no evidence for binary
orbital motion of this pulsar, at least with $P_{orb} < 250$ days
and with $a \cdot \sin{i} > 60$ ms.

\acknowledgments
This work was supported partly by grant INTAS 96-0154.

\end{document}